\documentclass{eas}
\usepackage{graphicx}
\begin{document}

\title{HVC and IVC gas in the halo of the Milky Way} 
\author{N. Ben Bekhti}\address{Argelander-Institut f\"{u}r Astronomie, Auf dem H\"{u}gel 71, 53121 Bonn, Germany}
\author{P. Richter}\sameaddress{1}
\author{T. Westmeier}\sameaddress{1}
\begin{abstract}
We combine CaII/NaI absorption and HI 21~cm emission line measurements to analyse the metal abundances, the distribution, the small-scale structure, and the physical conditions of intermediate- and high-velocity gas in the Galactic halo.

\end{abstract}
\maketitle
\section{Introduction}
Absorption and emission line measurements recently have demonstrated that the halo of the Milky Way comprises a population of neutral hydrogen clouds with low column densities. 
An efficient way to study metal abundances, small-scale structure and physical conditions in high- and intermediate-velocity gas moving through the halo of the Milky Way is to combine absorption line measurements in the direction of quasars with HI 21-cm radio observations (e.g., Richter {\em et al.\/} \cite{ref2001b}, Richter {\em et al.\/} \cite{ref2005}). 
Using UVES at the ESO VLT, we have detected weak high-velocity CaII and NaI absorption line systems in the direction of several quasars, suggesting the presence of filamentary or clumpy structures in the Galactic halo. Follow-up HI observations with the Effelsberg 100-m radio telescope have demonstrated that in many cases the CaII/NaI absorption is connected with neutral hydrogen gas.

\section{Results}

\begin{figure*}
\includegraphics[width=0.5\textwidth,clip]{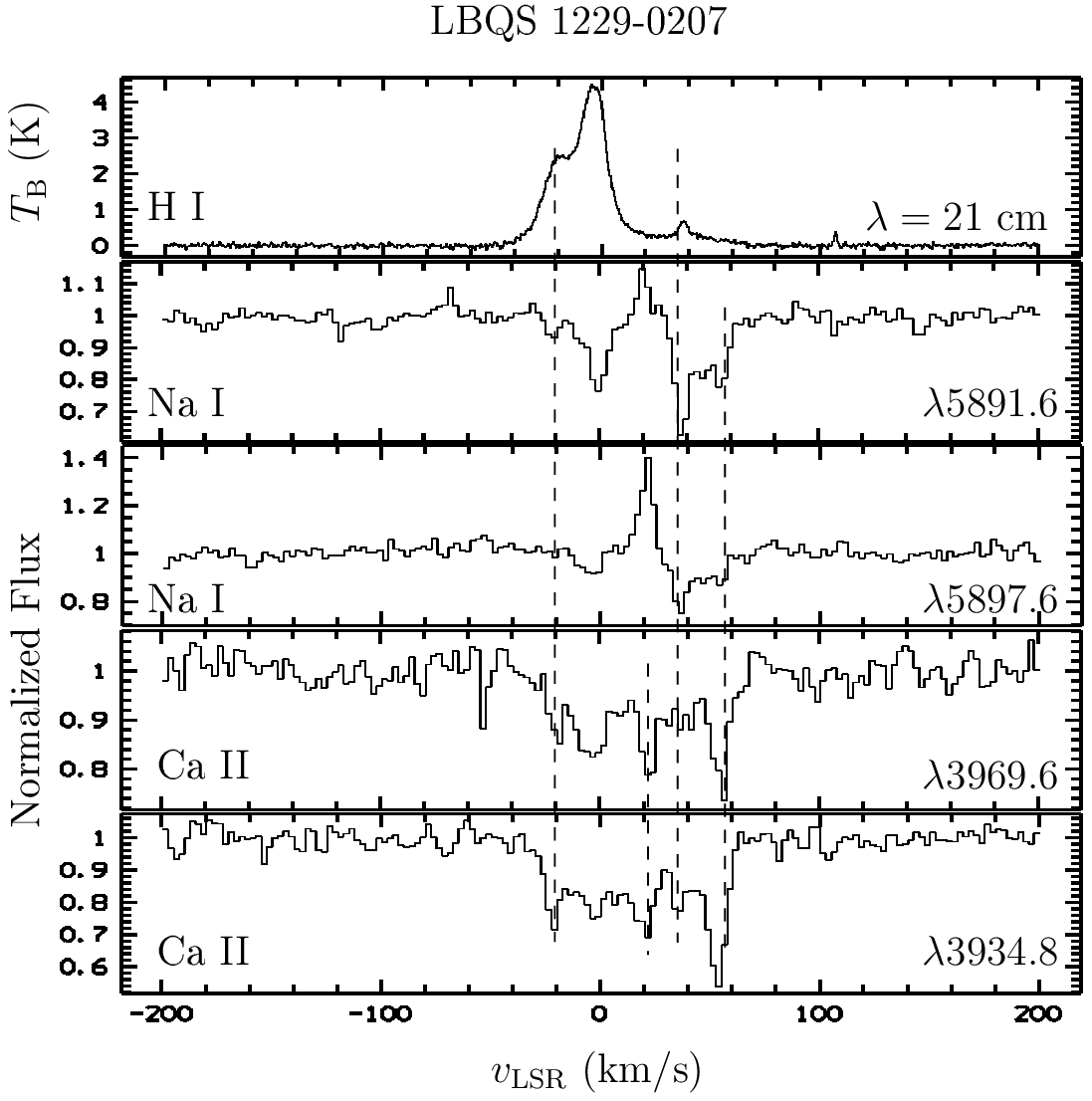}\hfill
\includegraphics[width=0.5\textwidth,clip]{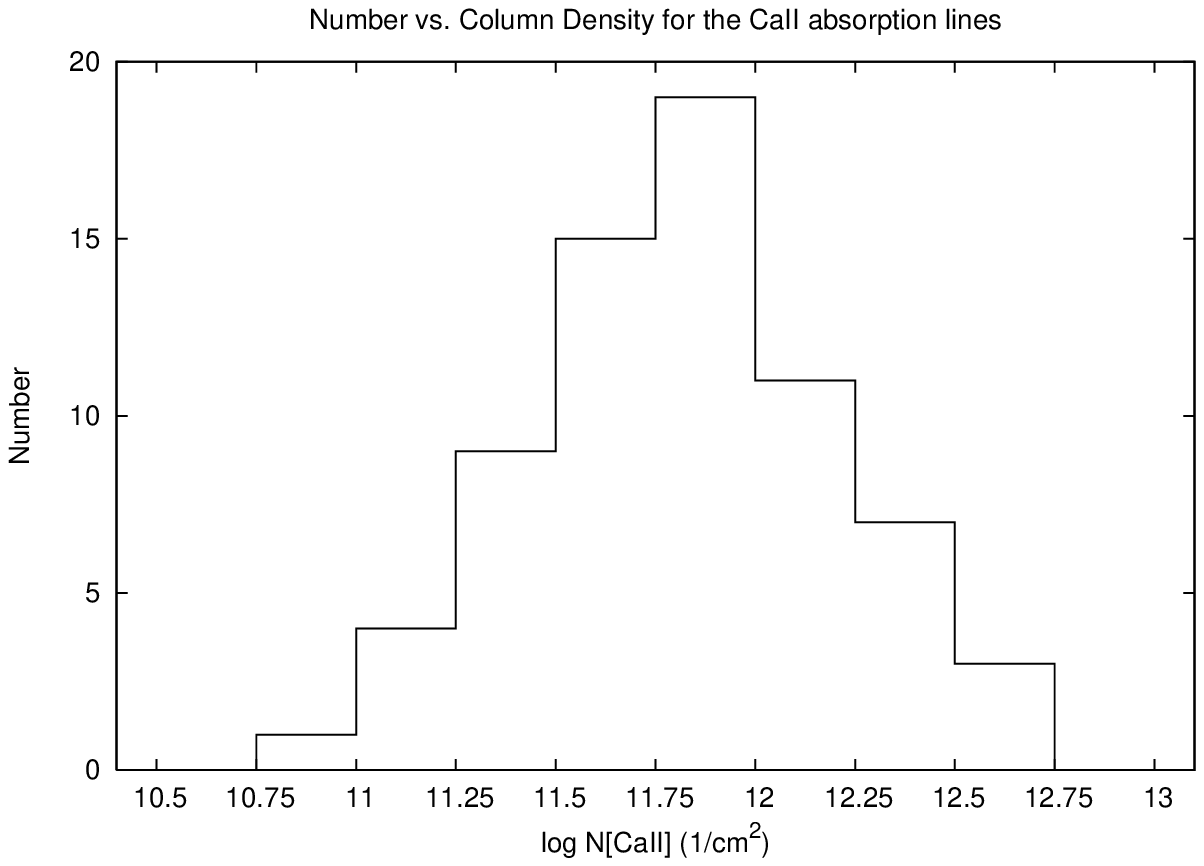}
\caption{\textbf{Left panel:} Absorption and emission spectra in the direction of the background quasar LBQS 1229--0207 observed with VLT/UVES and the 100-m telescope at Effelsberg. \textbf{Right panel:} Number of absorbing systems versus CaII column density.}
\label{fig_hvc_allskymap}
\end{figure*}

Our sample contains in total 95 sight lines observed with UVES. For 43 of them we have additional HI emission spectra obtained with the 100-m telescope at Effelsberg. The absorption profiles of CaII and NaI for one of the sight lines are plotted in the left panel of Fig.~\ref{fig_hvc_allskymap} on an LSR velocity scale together with the corresponding 21-cm emission spectrum.
The analysis of all spectra shows that in many cases the CaII/NaI absorption is connected with HI gas with column densities in the range of a few times $10^{18}~\mathrm{cm}^{-2}$ and $10^{20}~\mathrm{cm}^{-2}$. Our $3 \sigma$ HI column density detection limit is about $2 \cdot 10^{18} \; \mathrm{cm}^{-2}$ for the warm neutral medium. In a few cases the detected HI signals are below the detection limit of large HI surveys (e.g. LAB survey).
The measured HI line widths imply that for most of the sight lines the gas is relatively cold with temperatures $< 1000$~K. The directions and velocities of several of the clouds suggest a possible association with known HVC or IVC complexes. Most CaII/NaI absorbers show multiple intermediate- and high-velocity components, indicating the presence of filamentary or clumpy structures. The right panel of Fig.~\ref{fig_hvc_allskymap} shows the frequency of occurrence of the absorbing systems versus the CaII column density. There is a well-defined maximum between $\log N_\mathrm{CaII} \left[\mathrm{cm}^{-2}\right]=11.5$ and $\log N_\mathrm{CaII}  \left[\mathrm{cm}^{-2}\right]=12$, but completeness issues may apply. 

The fact that among 95~random lines of sight through the halo we observe 40~absorption systems with at least one intermediate- or high-velocity component suggests that the Milky Way halo is filled with low column density clouds. Their investigation with synthesis telescopes will allow us to determine their metallicities and, thus, to learn more about their origin. Studying the clouds will also allow us to draw conclusions about the absorption line systems widely observed in the halos of other galaxies (e.g., Bouch{\'e} {\em et al.\/} \cite{ref2006}) and the physical state of this gas.


\end{document}